# Space-Time-Modulated Wideband Radiation-Type Programmable Metasurface for Low Sidelobe Beamforming

Xudong Bai, *Member, IEEE*, Longpan Wang, Yuhua Chen, Xilong Lu, Fuli Zhang, Jingfeng Chen, *Member, IEEE*, Wen Chen, *Senior Member, IEEE*, and He-Xiu Xu, *Senior Member, IEEE*

*Abstract*—Programmable metasurfaces promise a great potential to construct low-cost phased array systems due to the capability of elaborate modulation over electromagnetic (EM) waves. However, they are in either reflective or transmissive mode, and usually possess a relatively high profile as a result of the external feed source. Besides, it is difficult to conduct multi-bit phase shift in metasurfaces, when comparing with conventional phased arrays. Here, we propose a strategy of space-time modulated wideband radiation-type programmable metasurface for low side-lobe beamforming. The wideband programmable metasurface avoids the space-feed external source required by its traditional counterpart, thus achieving a significant reduction of profile through integration of a high-efficiency microwave-fed excitation network and metasurface. Furthermore, through introducing space-time-modulated strategy, the high-accuracy amplitude-phase weight algorithm can also be synchronously carried out on the first harmonic component for low side-lobe beam-scanning. Most importantly, adaptive beamforming and generation of interference null can further be created after analyzing the harmonic component characteristics of received signals.

## I. INTRODUCTION

P HASED array antennas are extensively utilized in both military and civil areas, such as satellite communications, satellite navigation, spectrum reconnaissance and detection, and satellite remote sensing [1], [2]. The phased arrays achieve electronic beam steering through phase shifters, enabling rapid

Manuscript received July 1, 2024; revised 　　　, 2024; accepted 　　　, 2024. Date of publication 　　　; date of current version 　　. This work was supported by National Key Project under Grant 2020YFB1807700, National Natural Science Foundation of China under Grant 62071296, Shanghai Kewei Foundation under Grant 22JC1404000, Shanghai Aerospace Science and Technology Innovation Foundation under Grant F-202405-0018, and Basic Research Programs of Taicang under Grant TC2022JC16. *(Corresponding authors: He-Xiu Xu.)*
X. Bai, L. Wang, Y. Chen, X. Lu, and F. Zhang are with the College of Microelectronics, Northwestern Polytechnical University, Xi'an 710129, China (e-mail: baixudong@nwpu.edu.cn).
J. Chen and W. Chen are with the Department of Electronics Engineering, Shanghai Jiao Tong University, Shanghai 200240, China (e-mail: wenchen@sjtu.edu.cn).
H. -X. Xu is with the Air and Missile Defense College, Air Force Engineering University, Xi'an 710051, China (e-mail: hxxuellen@gmail.com).

scanning of tracking beams, the capability to alter beam shapes, and the dynamic synthesis of spatial power [3], [4]. However, the extensive requirement for numerous transceiver channels in existing phased array antenna systems not only results in significant cost and power consumption but also introduces complexity in design and manufacturing, greatly limiting their widespread application in aerospace systems, especially on small satellite platforms [5].

Programmable metasurfaces, the latest research branch of artificial electromagnetic (EM) metamaterials in recent years, are planar two-dimensional metamaterials typically composed of subwavelength-scale structural units arranged periodically or quasi-periodically [6], [7]. By integrating various active devices, programmable metasurfaces have shown hopeful applications in EM absorption, polarization control, and the design of novel microwave devices [8], [9], [10], [11]. Programmable metasurfaces achieve dynamic regulation of units by loading various reconfigurable devices and use bias circuits to control their dynamic encoding distribution, and thus realize flexible EM control of incident waves [12], [13], [14]. Novel phased array systems based on existing reflective or transmissive programmable metasurfaces, can significantly reduce the overall system cost and power consumption, by foregoing numerous T/R components required for phase shifting in traditional arrays [15], [16]. Existing programmable metasurfaces mainly include reflective and transmissive configurations associated respectively with the phase correction in scattering or transmission field of the meta-atom, both of which require an additional spatially excited feed source antenna [17], [18].

In 2014, the integration of digital coding characterization and field-programmable gate array (FPGA) was first introduced into the design of dynamic metasurfaces, thus constructing a reflective programmable metasurface system capable of real-time EM wave control by integrating switch diodes in unit structures [19]. Thereafter, a large-scale one-bit reflective programmable metasurface operating in the Ku-band was constructed, which was composed of five sub-arrays with a total of 1600 elements and later expanded to work in the X / Ku dual-band [20], [21]. To improve the channel capacity, the reflective programmable metasurfaces with independent control of dual polarizations were further studied, which could achieve parallel information transmission under two orthogonal channels by independently controlling the reflection phase of both *x*-polarized and *y*-polarized EM waves [22], [23]. Ultimately, a two-bit reflective programmable metasurface was designed by integrating two independently biased PIN diodes within the meta-atom to obtain four reconfigurable phase states, thus an adaptive intelligent wireless energy transmission system based on the digital metasurface was constructed [24].

Reflective programmable metasurfaces, while benefiting from simple meta-atom structures and low design complexity, are constrained by significant limitations such as poor phase stability and severe feed blockage [25]. Correspondingly, programmable metasurfaces on transmissive mode emerged, and the transmissive programmable metasurface capable of transforming incident linearly polarized EM waves into transmitted EM waves of any polarization was proposed [26]. Ulteriorly, an X-band transmissive programmable metasurface with high one-bit phase-tuning stability, was designed to stimulate dynamic vortex EM beams with high mode purity [27], and then a novel multi-tier computing network has been proposed for improving computing capability, based on the transmissive metasurface architecture [28]. Addressing the significant constraint of excessive overall profile height in conventional reflective or transmissive programmable metasurfaces, many in-depth studies have also been conducted. A novel form of folded programmable metasurface was developed, which halves the overall profile and enables wide-angle beam scanning by combining polarization-selective surfaces with reflective digital surface units featuring cross-polarization rotation capabilities [29]. Subsequently, this innovative concept was further extended to the design of dynamic multimode vortex waves with high mode purity [30]. Thereafter, a low-profile near-field transmissive programmable metasurface was constructed, which was excited by planar waves from high-efficiency waveguide arrays and further combined with high-dielectric constant dielectric lens to reduce the overall profile [31]. However, the dielectric lens also introduced energy loss and significantly increased the system complexity as well as the total weight.

To further expand the multidimensional control capabilities of the programmable metasurfaces over EM waves, numerous researches have been conducted. Since 2018, by combining the existing hardware architecture of programmable metasurfaces and the time-modulated array algorithms [32], [33], [34], the concept of space-time-coding digital metasurfaces was first proposed to achieve simultaneous spatial and frequency domain control of electromagnetic waves on a single digital metasurface [35], [36], [37]. Subsequently, this novel concept has been extended to Doppler frequency shift compensation and multi-user wireless communication systems, allowing independent control over multiple harmonic components for dynamic control of electromagnetic wave propagation direction and harmonic power distribution [38], [39], [40]. Ulteriorly, a space-time-coding anisotropic metasurface was constructed to dynamically stimulate arbitrary-polarization multi-beams, thus achieving joint control over polarization states and scattering patterns [41]. Until very recently, space-time-coding metasurfaces further demonstrate great potentials in direction-of-arrival (DoA) estimation and RCS manipulation [42], [43].

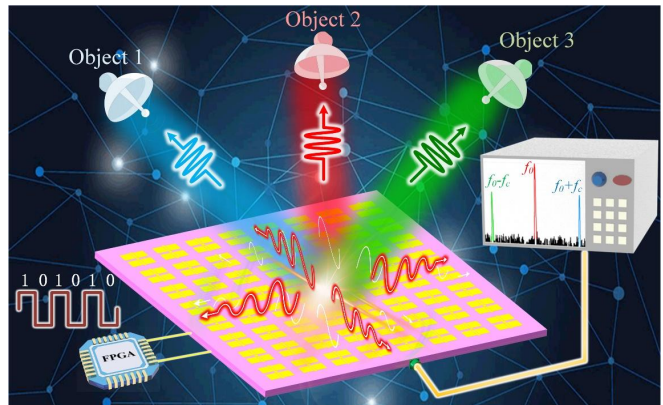

**Fig. 1.** Schematic diagram of the space-time-modulated wideband radiation-type programmable metasurface.

Overall, programmable metasurfaces have been applied to many research fields for effectively decreasing the cost of phased array system. However, existing researches on digital metasurfaces still fall short of achieving precise EM modulation for low-profile conformal applications, with major constraints as follows. Firstly, the profile of conventional programmable metasurfaces need further reduction since they possess a relatively high profile as a result of external feed source. Secondly, the phase resolution of metasurfaces also need further improvement since the phase shift accuracy is quite insufficient for ultra-low sidelobe beamforming in radar detection and secure communication applications; besides, it would be of great value to conduct joint control of both amplitude and phase response for the meta-atom [44].

In light of above challenges, a novel space-time modulated wideband radiation-type programmable metasurface is proposed, which could achieve a significant reduction of the overall profile through the integration design of a high-efficiency microwave excitation network and the supernatant functional metasurface configuration, along with the wideband one-bit phase tuning. Simultaneously, when further incorporating with space-time-coding modulation strategy, the precise amplitude-phase joint regulation can be ulteriorly achieved in the desired harmonic component, and the low sidelobe (SLL) dynamic beamforming has then been studied both numerically and experimentally. The rest of this paper is organized as follows. In Section II, the hardware architecture of the proposed metasurface is detailedly illustrated. Then in Section III, the temporal modulation scheme is numerically derived based on the space-time-modulated metasurface algorithm. In Section IV, the wideband radiation-type programmable metasurface is fabricated and experimented to conduct sufficient verification of our proposed design. Finally in Section V, the conclusion is provided.

## II. METASURFACE DESIGN AND SIMULATION

To overcome the limitations of high profile and leakage radiation present in conventional reflective or transmissive

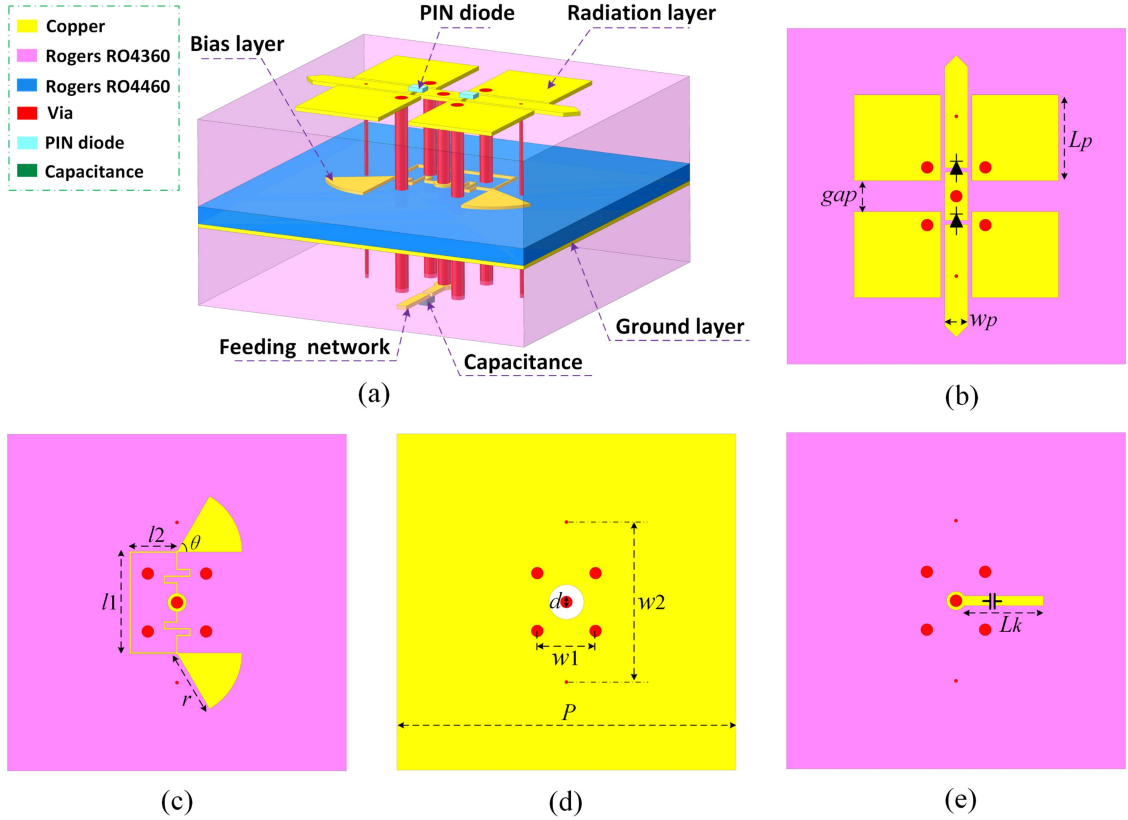

**Fig. 2.** Topology of the reconfigurable meta-atom. a) Three-dimensional illustration. b) Top radiation layer. c) Bias layer. d) Ground plane layer. e) Bottom feeding network. ( $P = 25\text{mm}$ , $Lp = 7.66\text{mm}$ , $gap = 2.45\text{mm}$ , $wp = 1.9\text{mm}$ , $r = 4.8\text{mm}$ , $\theta = 60°$ , $l_1 = 7.5\text{mm}$ , $l_2 = 3.5\text{mm}$ , $w_1 = 4.45\text{mm}$ , $w_2 = 8.86\text{mm}$ , $d = 0.9\text{mm}$ , $Lk = 6.35\text{mm}$ )

programmable metasurfaces, the topology of the radiation-type programmable metasurface is proposed by integrating the feeding network into the unified design of the digital phase-adjusting meta-atom, thus effectively reducing the overall profile height of the system. The overall schematic of the proposed space-time-modulated wideband radiation-type programmable metasurface is illustrated in Fig. 1, which fundamentally consists of periodic meta-atoms stimulated by the traveling wave propagating within the integrated feed network attached below the metasurface aperture. The EM wave is stimulated at the center of the underneath feed network layer with initial constant amplitude and same phase.

### A. Meta-atom design

To expand the effective operation bandwidth of the metasurface, the construction of the meta-atom could draw experience from design philosophy of the broadband magnetoelectric (ME) dipole antennas [45], [46], [47], which involves integration of complementary far-field characteristics of the electric and magnetic dipoles to achieve favorable radiation performance across a wide operating frequency range, thus significantly mitigating the backward radiation. It needs to be stressed that, the off-center feeding in conventional ME dipoles should be improved pertinently to fit within the wideband binary phase tuning for the proposed meta-atom.

The topology of the proposed wideband radiation-type programmable meta-atom is as shown in Fig. 2, which is composed of four metallic layers supported by two substrates layers (Rogers RO4360, dielectric constant of 6.15, loss tangent of 0.0038) and a bonding film layer (Rogers RO4460G2, dielectric constant of 6.15, loss tangent of 0.0045). The four metallic layers include the top radiation layer, the biasing layer, the metallic ground layer, and the feeding network. The top radiation layer includes the metasurface ME radiator, which consists of four quadrate metal patches along with four cylindrical metalized grounding via-holes. Furthermore, the appropriate coupling excitation method using T-shaped probe is employed to stimulate the metasurface radiator, thus obtaining a good impedance matching over a very wide frequency band. The T-shaped probe is composed of two parts and located between the metal patches. The first part of the T-shaped probe is a centre-fed metalized via-hole, which is also connected with the feeding network beneath the metallic ground. The second part of the T-shaped probe is a horizontal microstrip line, which is located between the metal patches with the same height, in order to couple the RF signal into the metasurface radiator.

To obtain the stable binary phase quantization difference within a broad operation bandwidth, the proposed radiation-type meta-atom employs the current-inversion phase-shifting method, which is achieved by integrating two PIN diodes anti-symmetrically configured on the horizontal part of the T-shaped coupling probe, thus utilizing the spatial change in the excitation current direction for a wideband reverse-phase modulation. In practical designs, it is necessary to extract the equivalent lumped components of the PIN diodes for the two coding states, and the PIN diode, MACOM MADP- 000907-14020 [48], is carefully selected to obtain low insertion losses within a wide operation band. When the PIN diode is switched by positive biasing, the equivalent circuits of series resistor $R_{ON}$ = 5.2 Ω and inductor $L_{ON}$ = 30 pH are configured; while switched by negative biasing, the equivalent capacitance $C_{OFF}$ = 0.025 pF and inductor $L_{OFF}$ = 30 pH are selected.

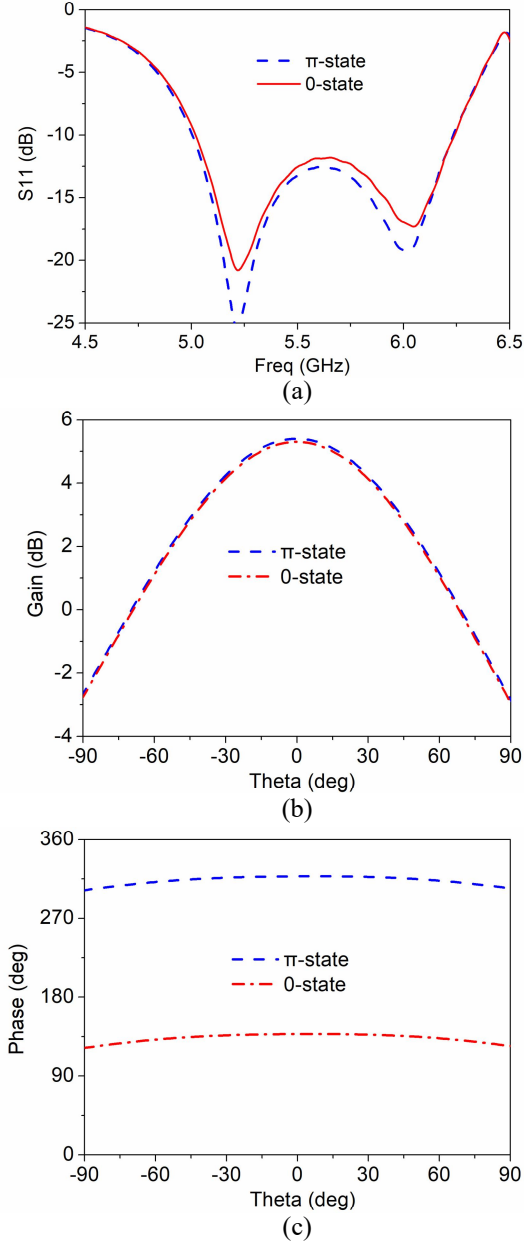

**Fig. 3.** Simulated property of the radiation-type programmable metasurface for both π / 0 states. a) Reflection coefficient $S_{11}$. b) Radiation pattern at 5.5GHz. c) Radiation phase at 5.5 GHz.

To further achieve real-time dynamic control over the coding states of the meta-atom, the biasing control layer needs to be integrated into the unit configuration for managing the operation states of the PIN diodes. The integrated configuration of the proposed radiation-type meta-atom inevitably causes strong coupling interference between the high-frequency EM waves and the high-speed DC control signals, since the biasing control circuits could significantly affect the actual radio frequency (RF) performance. Therefore, designing an effective biasing control circuit is another key point in the design of the radiation-type programmable metasurface. To minimize the impact of the biasing circuit on the unit performance, the biasing layer is designed with narrow-linewidth high-resistance strip-lines and placed very close to the metallic ground. Furthermore, symmetrical zigzag-line inductors and fan-shaped distributed capacitors are integrated into the biasing control circuits to choke the high-frequency signal. Afterwards, the DC-blocking capacitor should also be integrated in the bottom feeding network in order to prevent short-circuit between the biasing control lines of different meta-atoms.

The simulation verifications of the proposed radiation-type programmable meta-atom are further carried out and the results are provided in Fig. 3. The meta-atom could exhibit a wide operation bandwidth for reflection coefficients $S_{11}$ < −10dB, covering the frequency from 5.82 GHz to 6.25 GHz for both programmable states, with a relative operation bandwidth over 20% and an overall profile height of only about one-tenth of the free-space wavelength. The simulated far-field radiation patterns for both π / 0 states of the designed meta-atom at the central frequency 5.5 GHz are also provided in Fig. 3(b), and the meta-atom could obtain the realized gains over 5.3 dB for both states with the 3-dB beamwidth over 100°. The simulated radiation phase pattern of the meta-atom for both states are also plotted in Fig. 3(c), and the phase difference of the binary programming states could stabilize at around 180° of little variation within the whole operation band. The above simulation results verify adequately the effectiveness of our proposed radiation-type meta-atom design.

### B. Metasurface array design

Based on the above meta-atom, the integrated metasurface array is further designed, and a highly integrated architecture with the metasurface radiators, the microwave excitation networks, and the DC bias circuits is constructed to replace the spatial feeding method in conventional programmable metasurfaces, thus achieving a low-profile and highly efficiently integrated metasurface system. To form a comprehensive design method for dynamic electromagnetic control of radiation-type programmable metasurface, the efficient transmission coupling mechanisms between the transmission network and the metasurface radiators should be lucubrated and exploited, based on the multilayer RF circuit integration method.

The programmable metasurface is well-designed and the overall metasurface configuration is presented in Fig. 4, which is composed of 8 × 8 units along with 128 PIN diodes. The power divider network used for active radiation-type unit feeding of the programmable metasurface system can take various forms, such as microstrip networks, waveguide power divider networks, and substrate integrated waveguide power divider networks. For practical design, by compromising the system complexity, the profile height, and the overall manufacturing cost, a parallel equal-split microstrip network is preferably adopted to stimulate the initial equal amplitude and same phase for the meta-atoms, thereby effectively reducing the complexity of the programmable metasurface algorithms.

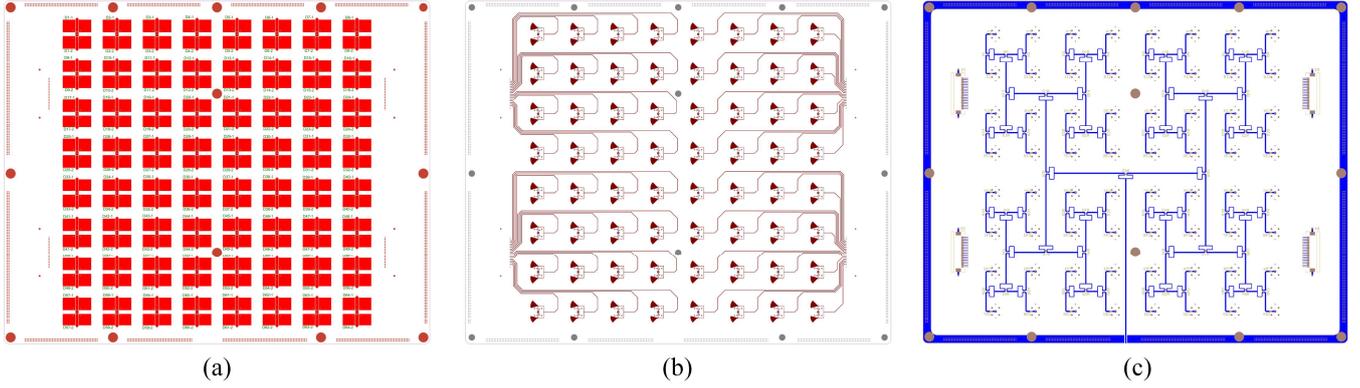

**Fig. 4.** Configuration of the proposed metasurface with 8 × 8 units. a) Top layer with the radiation metasurface patches. b) Bias layer with the bias network layout. c) Bottom microwave feed network layer.

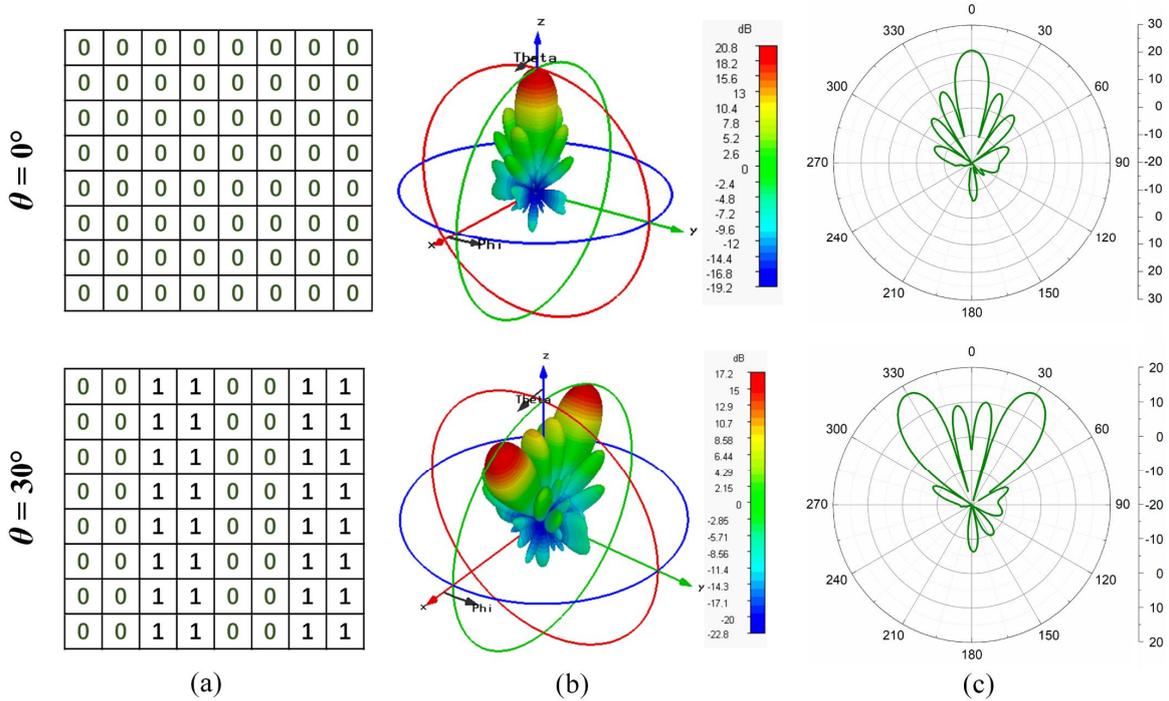

**Fig. 5.** Property of the proposed metasurface for two scanning angles $\theta = 0°$ and $\theta = 30°$ in H-plane at 5.5 GHz. a) Code distribution. b) Simulated 3D radiation pattern. c) Simulated 2D radiation pattern.

Subsequently, a high degree of integration for the metasurface radiators, microwave excitation networks, and DC bias circuits is achieved through multilayer RF circuit integration. The bias control layer shown in Fig. 4(b) is placed very close to the metal ground plane, with a mirror-symmetric distribution to effectively reduce the interference between bias lines for different meta-atoms. 64 capacitances are also integrated into the high-frequence feeding network to ensure the independence of the DC control lines between different meta-atoms, thus effectively preventing the risk of short circuits between bias control lines caused by RF circuits. By implementing parallel dynamic high-speed control through FPGA, different bias voltages can be applied to control the ON and OFF states of PIN diodes on each unit of the metasurface system, thus the radiation beams of the digital meta-atoms can be stimulated to implement different coding states. Furthermore, according to the actual functional requirements, scanning beams or shaped beams with different directions can be stimulated, through modulating the coding distribution on the radiation-type programmable metasurface.

The parallel microstrip network designed to feed the supernatant patches is shown in Fig. 4(c). Wideband Wilkins power dividers are employed and the electromagnetic wave could propagate outward from the center through the integrated feed network with initial constant amplitude and same phase. Through controlling the quantized aperture coding distributions via the FPGA, each unit could implement the binary phase coded modulation based on the feeding power, and the dynamic radiation patterns can thus be created for the proposed radiation-type programmable metasurface.

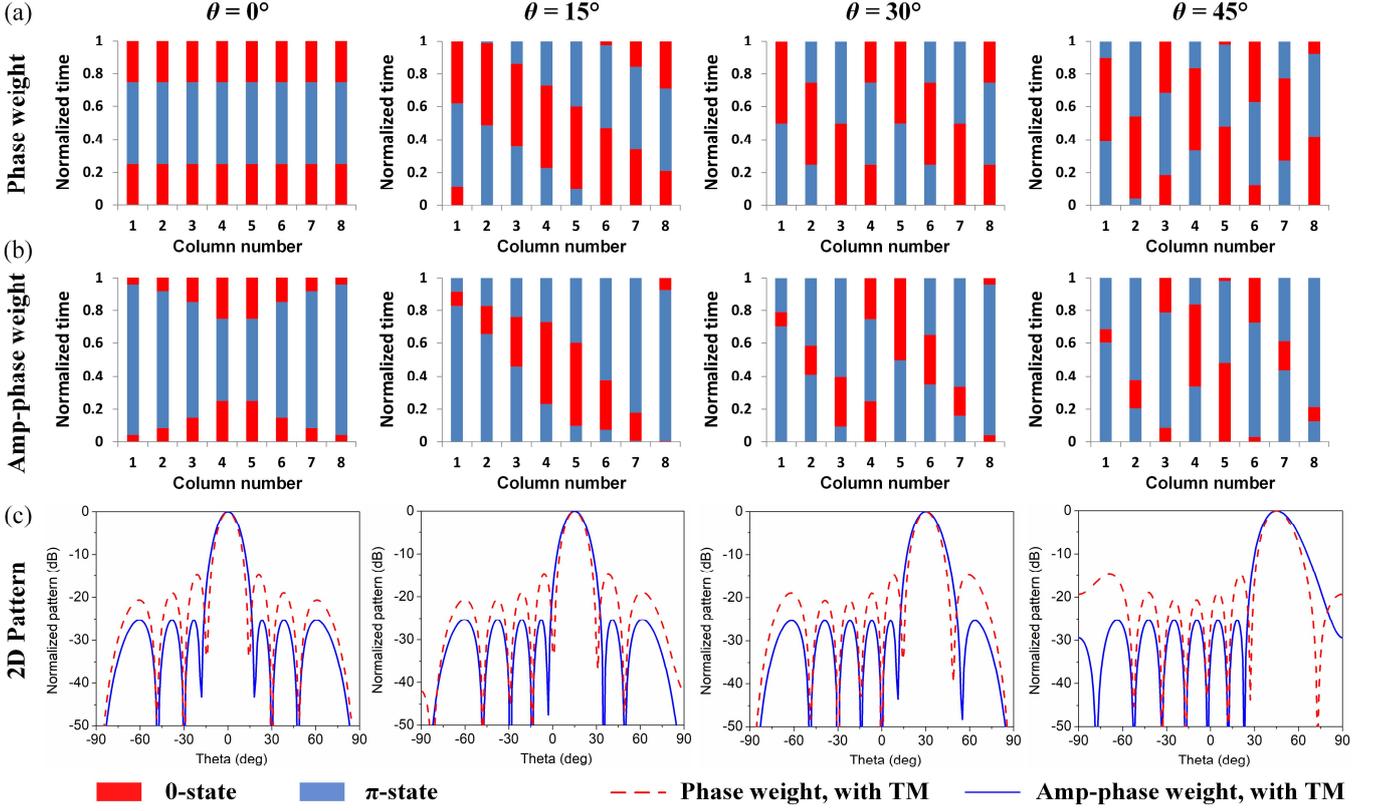

**Fig. 6.** Time modulation coding sequences and the corresponding 2D radiation patterns for different scanning angles $\theta = 0°$, $\theta = 15°$, $\theta = 30°$ and $\theta = 45°$. a) Time modulation coding sequences for phase weight. b) Time modulation coding sequences for both amplitude and phase weight. c) Radiation patterns for the $+1^{st}$ harmonic component for two conditions.

Fig. 5 gives the representative examples for generating directional single-beam and dual-beam radiations through the radiation-type programmable metasurface, and the synthesized aperture distributions along with the homologous simulated far-field radiation patterns are also provided. It is observed that, good-shaped directional pencil beam can be created for a consistent metasurface code distribution, while the dual-beam radiation pattern would be generated through the alternate distribution of the two programmable states.

## III. SPACE-TIME-MODULATED METASURFACE

Based on the research design of the 1-bit radiation-type programmable metasurface, the space-time-modulated strategy is further combined with the metasurface hardware through the high-speed modulation of the real-time periodic coding sequences. By utilizing an FPGA to code the binary phase states of the meta-atom with time modulation sequences periodically, harmonic components are thus generated. Through introducing space-time-modulated strategy, the high-accuracy amplitude-phase weight algorithm can then be synchronously carried out on the first harmonic component, leading to further optimization of the amplitude and phase distribution of electromagnetic waves in three-dimensional space, and the low-sidelobe dynamic beam scanning along with high precision beam steering can thus be obtained.

From the Fourier series expansion properties of periodic functions, it is known that the energy carried by a RF signal will be redistributed among the fundamental and harmonic components after periodic phase modulation [44], [49], and the periodic modulation signal loaded on the $n^{th}$ unit of the metasurface can be expressed as:

$$U_n(t) = \sum_{m=-\infty}^{\infty} g_n(t - mT_p) = \sum_{k=-\infty}^{\infty} \alpha_n^k e^{jkW_p t}, \quad (1)$$

where $T_p$ is the modulation period, $W_p = 2\pi / T_p$, and $g_n(t)$ is indicated as

$$g_n(t) = \begin{cases} +1, \tau_{on,n} < t \leq \tau_{off,n} \\ -1, 0 < t \leq \tau_{on,n} \text{ or } \tau_{off,n} < t \leq T_p \end{cases}. \quad (2)$$

where $\tau_{on,n}$ and $\tau_{off,n}$ is the opening and closing moment of the "$\pi$" phase state.

Then the $k^{th}$ harmonic component can be deduced as

$$\begin{aligned} \alpha_n^k &= \frac{1}{T_p} \int_0^{T_p} g_n(t) e^{-jkW_p t} dt \\ &= \begin{cases} 0, k = 0 \\ \frac{j}{2\pi k}(1 - e^{-jk\pi})\left(e^{-jkW_p\tau_{off,n}} - e^{-jkW_p\tau_{on,n}}\right), k \neq 0 \end{cases} \end{aligned} \quad (3)$$

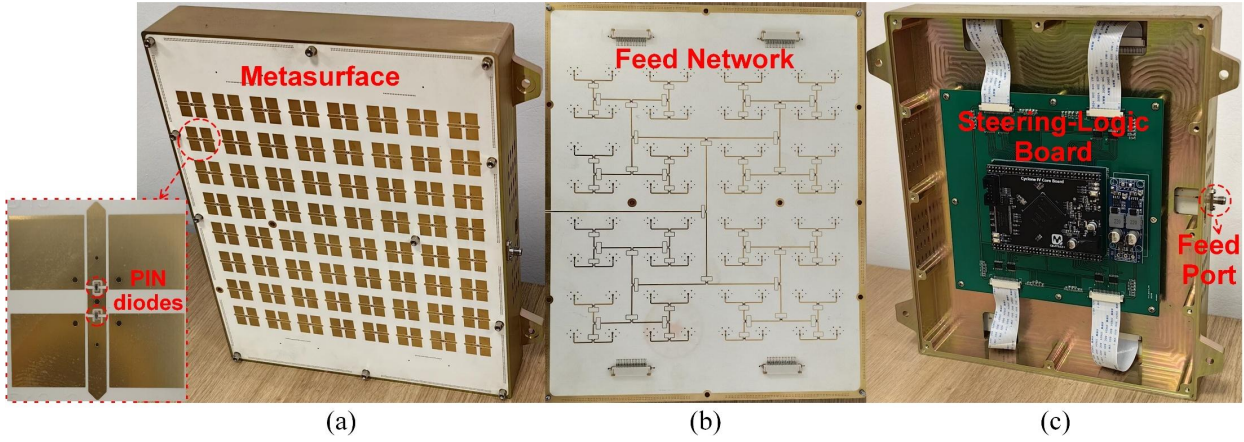

**Fig. 7.** Fabricated prototype of the wideband radiation-type programmable metasurface. a) Front metasurface array.  b) Microwave feed network. (c) The steering-logic board.

thus the $+1^{st}$ harmonic component can be derived as

$$\alpha_n^1 = \frac{2\sin\left[W_p(\tau_{off,n} - \tau_{on,n})\,/\,2\right]}{\pi} e^{-jW_p(\tau_{off,n} + \tau_{on,n})/2} . \quad (4)$$

Based on above analysis, we know that the energy of the RF signal could be modulated to the harmonic components by controlling the programmable metasurface system through using the FPGA to switch the meta-atoms between two phase states ($0\,/\,\pi$), respectively, during a modulation period.

According to (4), the power distributed for the $+1^{st}$ harmonic component would be restructured accordingly, when the ratio of the timing lengths for the two coding phase states is varying. Therefore, the amplitude weight of the meta-atom can be realized by assigning the ratio of the timing lengths for the two coding phase states in a modulation period. Also based on (4), the phase weight of the digital meta-atom can be obtained by modulating the time delay of the FPGA control signal. If the control signal on the $m^{th}$ unit is $S_c(t)$, and the control signal on the $n^{th}$ unit is $S_c(t-u)$, then the phase difference of the first harmonic component generated on the $m^{th}$ and $n^{th}$ units would be $2\pi u\,/\,T_p$.

Taking the $M \times N$ radiation-type programmable metasurface as an example, assuming that the initial amplitude and phase of the $(m,n)^{th}$ unit are $a_{m,n}$ and $\phi_{m,n}$, respectively. In order to achieve ultra-low sidelobe beam scanning, we can combine the equiphase surface method with the Chebyshev array synthesis method to calculate the amplitude weight $\gamma_{m,n}$ and phase weight $\phi_{m,n}$ for each unit on the metasurface. Combining the initial amplitude distribution of the units on the radiation-type metasurface, the amplitude of the $(m,n)^{th}$ unit needs to be compensated as

$$\beta_{m,n} = \gamma_{m,n}\,/\,a_{m,n}. \quad (5)$$

For the $(m,n)^{th}$ radiation-type meta-atom, within a modulation period $T_p$, setting the phase to 0-state for time $\left(0, \tau_{m,n}\right]$ and $\pi$-state for time $\left(\tau_{m,n}, T_p\right]$, the first harmonic component generated would be given by

$$A_{m,n} = \frac{2}{\pi}\sin\frac{W_p\tau_{m,n}}{2}e^{-j\frac{W_p\tau_{m,n}}{2}} . \quad (6)$$

It can also be obtained that the time $\tau_{m,n}$, when the coding phase state switches from 0 to $\pi$, can be expressed as

$$\tau_{m,n} = \frac{2}{W_p}\arcsin\frac{\pi\beta'_{m,n}}{2}, \quad (7)$$

where $\beta'_{m,n}$ is a normalized parameter with respect to $\beta_{m,n}$. Noted that, since the additional phase shift $-W_p\tau_{m,n}\,/\,2$ would be introduced by the periodic modulation; therefore, when introducing the control timing delay $u_{m,n}$ to adjust the phase shift for the positive first harmonic component, the following equation should be satisfied,

$$-W_p u_{m,n} - W_p\tau_{m,n}\,/\,2 - \varphi_{m,n} = \phi_{m,n} + 2k \quad (8)$$

Then, the time delay $u_{m,n}$ of the control timing corresponding to the $(m,n)^{th}$ meta-atom can be calculated with respect to the reference sequence.

Based on above theoretical analysis, numerical calculations are furtherly performed to verify the time-modulated high-precision amplitude phase control method. Taking previously designed 8×8 unit radiation-type programmable metasurface as an example, the spacing between neighboring meta-atoms is $\lambda\,/\,2$, and the system clock of FPGA is 64 MHz with modulation frequency 1MHz. Its minimum phase step can be equivalent to the phase shift accuracy of a 6-bit digital phase shifter.

By combining the equiphase surface method, Chebyshev array synthesis method, and the space-time-modulated method, the directional pattern of the positive first harmonic component generated by space-time modulation is synthesized. Through simulation, the two-dimensional direction maps are obtained when the scanning beam points to different angles. And the two cases in time modulation of phase weight and amplitude-phase weight are compared at the same time. The detailed time modulation coding sequences and the corresponding radiation patterns for different scanning angles $\theta = 0°$, 15°, 30° and 45°, are provided in Fig. 6. It is observed that, the beam pointing accuracy of the metasurface could be greatly improved after employing the time-modulated amplitude-phase weight algorithm, and the sidelobe levels would be reduced dramatically when comparing to that without time modulation. Besides, there could also be remarkable SLL reductions for the metasurface with both amplitude and phase weight, when comparing to that with only phase weight.

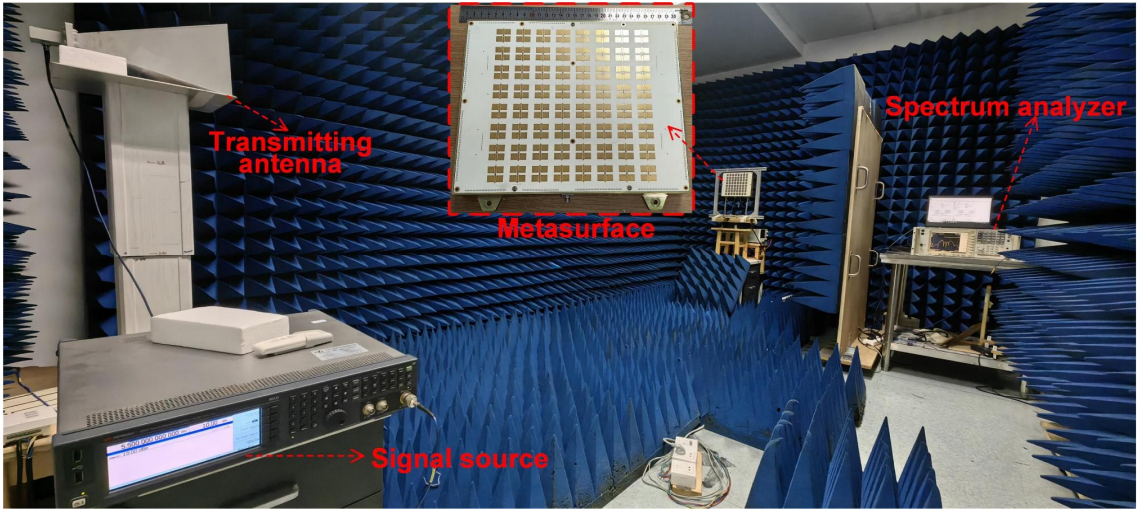

**Fig. 8.** Experimental environment and setup in the microwave anechoic chamber for the wideband radiation-type programmable metasurface.

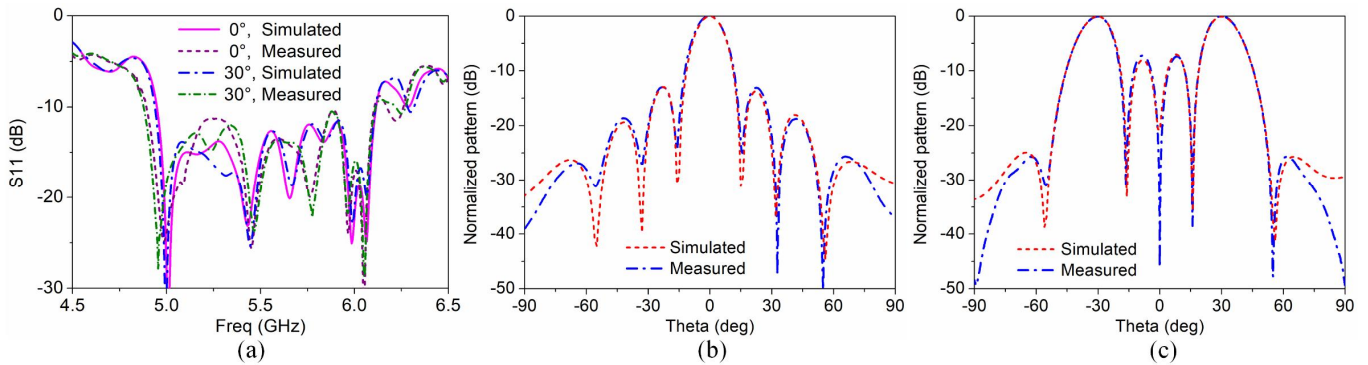

**Fig. 9.** Experimental and simulated results of the proposed radiation-type programmable metasurface without time modulation in H-plane. a) Reflection coefficient S11 for scanning angles $\theta = 0°$ and $\theta = 30°$. b) 2D radiation patterns for scanning angles $\theta = 0°$ at 5.5 GHz. c) 2D radiation patterns for scanning angles $\theta = 30°$ at 5.5 GHz.

## IV. EXPERIMENTAL VERIFICATION

To verify the actual operation performance of our proposed space-time-modulated wideband radiation-type programmable metasurface for dynamic low sidelobe beamforming, the metasurface prototype with $8 \times 8$ units was fabricated with the overall dimension of $266mm \times 216mm$, as shown in Fig. 7. The actual fabrication of the radiation-type metasurface utilizes a conventional multi-layer printed circuit board manufacturing process, and a total of 128 PIN diodes are soldered across the entire metasurface array, where each meta-atom is individually modulated by the steering-logic board through 64-channel independent bias control lines. The metasurface prototype is placed on the upper construction of a specially-constructed metal frame, while the steering-logic board is installed below the metal frame so as to effectively reduce the impact of the bias circuits on the metasurface radiation. The metasurface is connected with the steering-logic board through four flexible flat cables, and the high-frequency SMA connectors are also mounted on the metal frame for efficient excitation to the microwave feed network. The steering-logic board is powered by a 12V-voltage external input power source supply, which could provide positive or negative bias voltages for each meta-atom. The host processing system employs a FPGA, to adjust the independent 64-channel control signals within the assigned clock according to the optimal quantization code distribution, thus achieving efficient wideband dynamic beam modulation. For either phase state of the meta-atom, the designed positive or negative drive current of 10mA is provided on each bias line, with an additional DC bias voltage of ±1.2V. Based on the above parameters, the total power consumption of the whole system can be estimated to be within 1.0 W.

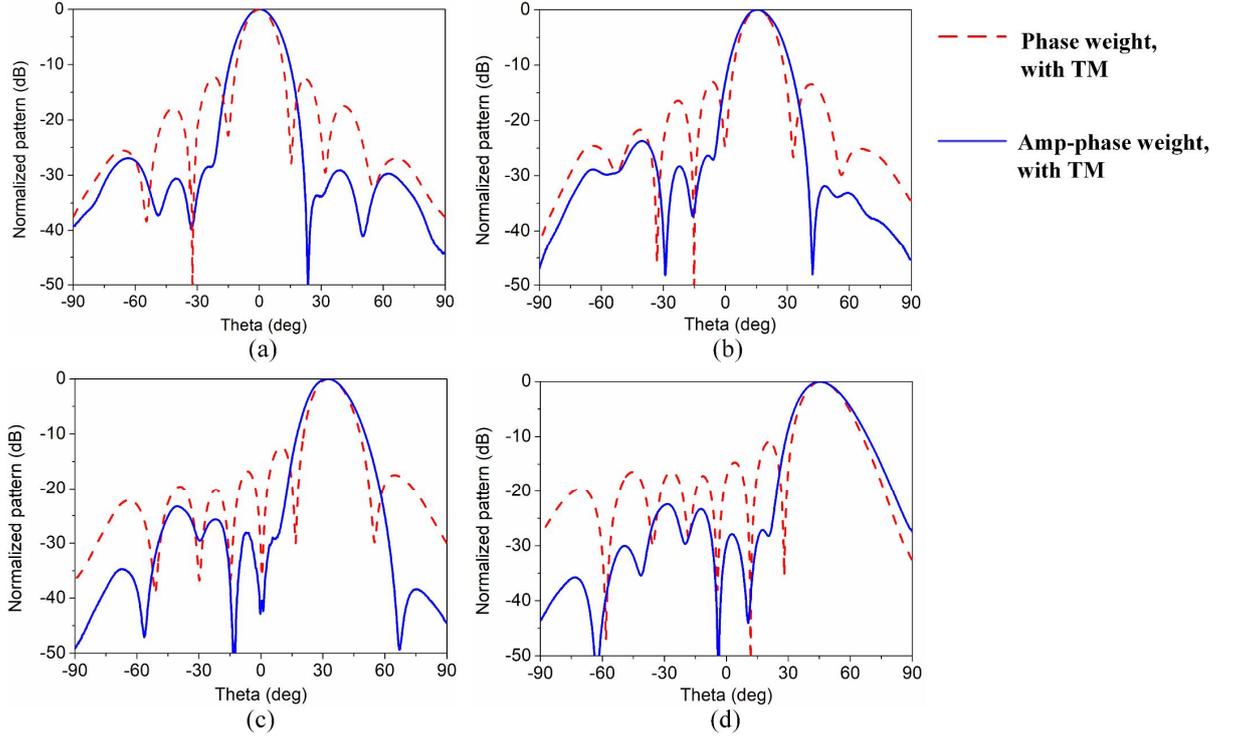

**Fig. 10.** Experimental radiation patterns for different scanning angles with only phase weight or amp-phase weight through time modulation in H-plane at 5.501 GHz. a) $\theta = 0°$. b) $\theta = 15°$. c) $\theta = 30°$. d) $\theta = 45°$.

The metasurface far-field test is conducted in the microwave anechoic chamber to measure its radiation characteristics, including array gain, radiation pattern and system beamwidth. The actual experimental environment along with far-field measurement setup is also presented as shown in Fig. 8, and the prototype of the programmable metasurface along with the designed metal framework is fixed to the test turntable. The rotational speed of the test turntable is controlled by the host computer, and set to synchronize with the spectrum, so as to acquire the accurate harmonic pattern data. A standard horn antenna is connected to a signal source as the transmitting antenna. The radiation-type programmable metasurface to be tested is connected to the spectrum analyser as the receiving antenna, and its radiation power patterns can then be determined by comparing with the measured receiving level of the standard horn antenna. It should be noted that, the conventional method by using a vector network analyzer to acquire the radiation patterns would be no longer applicable, owing to the fact that the operation frequency of the signals would be shifted from the fundamental component to the corresponding harmonic components through the introduction of the space-time modulation algorithms.

Fig. 9 presents the measured and simulated results for the proposed radiation-type programmable metasurface without time modulation for two scanning angles $\theta = 0°$ and $\theta = 30°$. It is observed that, the comparison between the experimental and simulated results generally presents the same variation tendency, and the measured reflection coefficients cover the effective frequency band from 4.892 GHz to 6.105 GHz for $S_{11} < -10\text{dB}$, acquiring a wide fractional operation bandwidth of about 22%. Fig. 9(b) presents the simulated and measured 2D far-field normalized patterns for $\theta = 0°$ at 5.5 GHz, and the relevant maximum measured gain is 20.6 dBi with the aperture efficiency of about 68%, and the experimental results agree well with the simulation prediction. The normalized measured and simulated patterns for $\theta = 30°$ are also provided as shown in Fig. 9(c), and the mirrored beams are demonstrated for the main lobes of the metasurface, thus acquiring a dual-beam radiation pattern performance.

Based on the space-time-modulated metasurface theory, both amplitude and phase weight on the meta-atom could be translated as the time modulation coding sequences, and the normalized 2D radiation patterns with only-phase weight or with both amplitude and phase weight are measured and plotted, as shown in Fig. 10. The operation frequency of the transmitting horn antenna is set at 5.5 GHz, and the modulation frequency of the FPGA is assigned as 1 MHz, thus the +1st harmonic component would be located at 5.501 GHz. It can be seen that, in the four scanning directions of $\theta = 0°$, 15°, 30° and 45°, the proposed metasurface could obtain well-shaped directional pencil beams with superhigh pointing accuracy. The main lobes of the metasurface would cast off the mirrored beams shown in Fig. 9(c) and acquire the measured SLLs lower than −13 dB through introducing the phase weight of time-modulated strategy. As for the space-time-modulated strategy with both amplitude and phase weight, the beamwidth for the main lobes of the scanning beams would be broaden, and accordingly, the measured SLLs would be lower than −23 dB for all the scanning beams, with an overall remarkable SLL reduction of about −10 dB.

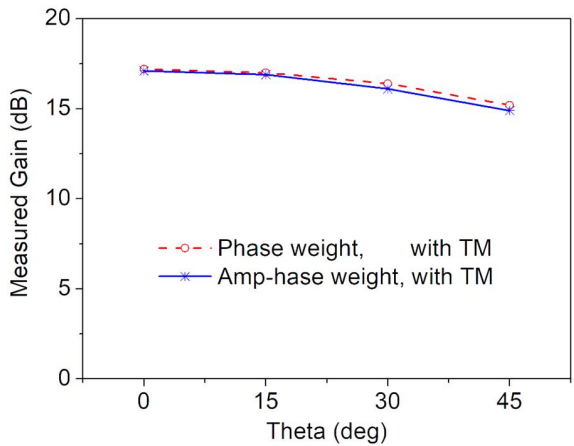

**Fig. 11.** Measured gains for different scanning angles with time modulation at 5.501 GHz.

Besides, the beamwidth of the main lobes would also have broadened with the increasing of the scanning angle. Fig.11 further presents the experimental array gains for different scanning angles of the proposed metasurface with time modulation, and there could be a slight decrease in the metasurface gains with both amplitude and phase weight, when comparing to that with only phase weight, indicating that the energy of the sidelobes have been effectively concentrated on the main lobe.

## V. CONCLUSION

In summary, we have presented a space-time-modulated wideband radiation-type programmable metasurface for low side-lobe beamforming. The proposed radiation-type metasurface could avoid the space-feed external source required by the traditional programmable metasurface, thus achieving a significant reduction of the profile through integration of a high-efficiency microwave-fed excitation network and the supernatant metasurface radiators. Based on numerical simulations of the constructed metasurface model, an 8×8-unit proof-of-concept programmable metasurface was fabricated and measured, and the effectiveness of proposed radiation-type strategy was verified. Ulteriorly, through introducing space-time-modulated strategy, the high-accuracy amplitude-phase weight algorithm can also be synchronously carried out on the first harmonic component. The radiation-type programmable metasurface could obtain a wide fractional operation bandwidth of about 22%, a measured gain of 20.6 dBi with an aperture efficiency over 68%, and reduced SLLs of −23 dB after introducing the amplitude and phase weight by the time-modulation coding sequences. Our strategy sets up a solid platform to design programmable metasurface with low profile for prospective applications in radar detection and low-sidelobe secure communication. Most importantly, adaptive beamforming and generation of interference null can further be created after analyzing the harmonic component characteristics of received signals.

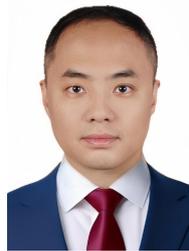

**XUDONG BAI** (Member, IEEE) received the B.S. degree, the M.S. degree and the Ph.D. degree in electronic science and technology from Shanghai Jiao Tong University (SJTU), Shanghai, China, in 2009, 2012 and 2016, respectively.

From 2016 to 2021, he was a senior engineer with the China Aerospace Science and Technology Corporation (CASC), China. Since 2022, he has been an Associate Professor with the School of Microelectronics, Northwestern Polytechnical University (NWPU), China. He has authored or coauthored more than 80 papers and holds more than 16 patents in antenna and metamaterial technologies. His current research interests include electromagnetic metasurfaces, low-cost phased arrays, and OAM-EM wave propagation and antenna design.

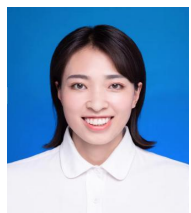

**LONGPAN WANG** received the B.S. degree in communication engineering from Xiangnan College (XNU), China, and the M.S. degree in electronic science and technology from Yantai University (YTU), China, in 2015 and 2019, respectively, and she is currently pursuing the Ph.D. degree in integrated circuit science and engineering from Northwestern Polytechnical University (NWPU), China, in 2022 until now.

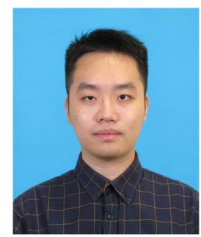

**YUHUA CHEN** received the B.S. degree in microelectronic science and engineering from Northwestern Polytechnical University (NWPU), Xi'an, China, in 2024 and is pursuing the M.S. degree in electronic science and technology from Northwestern Polytechnical University (NWPU), Xi'an, China, until now. His current research interests include electromagnetic metasurfaces, low-profile phased arrays, and OAM-EM wave propagation and antenna design.

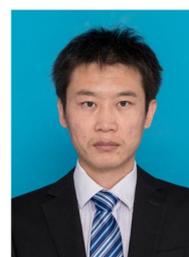

**XILONG LU** was born in Gansu, China, in 1987. He received the B.S. degree in physics from Lanzhou University, Lanzhou, China, in 2010, and the Ph.D. degree from the State Key Laboratory of Low-Dimensional Quantum Physics, Department of Physics, Tsinghua University, Beijing, China, in 2015.

From 2015 to 2021, he was with the Nanjing Research Institute of Electronics Technology, Nanjing, China, as a microwave and millimeter-wave integrated circuits senior engineer for active phased array radar applications. Since 2022, he has been an Associate Professor with the School of Microelectronics, Northwestern Polytechnical University (NWPU), Xi'an, China. His research interests include design and analysis of phased array radar, active and passive circuit, three- dimensional integration and package, 3D-SIP, HTCC / LTCC multichip module (MCM) technology, transceiver (TR) module, high-temperature super- conducting filter and system applications.

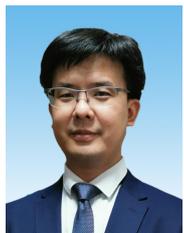

**Fuli Zhang** received the B.S. degree in materials science and engineering, the M.S. degree and the Ph.D. degree in optical engineering from Northwestern Polytechnical University, Xi'an, China, in 2003, 2006 and 2009, respectively.

He is currently a professor at Northwestern Polytechnical University. His research interests include micro and nano optics, digital metasurface and electromagnetic optical properties in RF devices. In recent years, he has published more than 70 papers in Advanced Materials, Physical Review Letters, Physical Review X and other important journals, and includes 2 ESI highly cited papers. He has more than 2,700 SCI citations and an H-index of 28, making him one of the World's Top 2% Scientists 2021. He also served as editorial board member of *Frontiers in Materials*, guest deputy editor of *Frontier in Physics*, the editor of *International Journal of Antenna Propagation*, reviewers for international journals such as *Nature Communications* and *Laser & Photonics Reviews*. He was invited to serve as the branch chairman of China Metamaterials Conference, IcAUMS, META and other conferences.

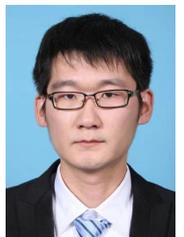

**JINGFENG CHEN** (Member, IEEE) received the B.S. degree in electronics and information engineering and the M.S. degree in signal and information processing from the Nanjing University of Information Science and Technology, Nanjing, China, in 2009 and 2012, respectively, and the Ph.D. degree in electromagnetic and microwave technology from Shanghai Jiao Tong University, Shanghai, China, in 2018. From 2019 to 2022, he was a Post-Doctoral Researcher with the Department of Electronic Engineering, Shanghai Jiao Tong University. In 2022, he joined Shanghai Jiao Tong University as an Associate Professor. His research interests include antenna arrays, unconventional array design, digital beamforming, and direction of arrival (DOA) estimation.

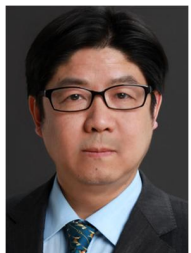

**WEN CHEN** (M'03–SM'11) received BS and MS from Wuhan University, China in 1990 and 1993 respectively, and PhD from University of Electrocommunications, Japan in 1999. He is now a tenured Professor with the Department of Electronic Engineering, Shanghai Jiao Tong University, China, where he is the director of Broadband Access Network Laboratory. He is a fellow of Chinese Institute of Electronics and the distinguished lecturers of IEEE Communications Society and IEEE Vehicular Technology Society. He is the Shanghai Chapter Chair of IEEE Vehicular Technology Society, a vice president of Shanghai Institute of Electronics, Editors of *IEEE Transactions on Wireless Communications*, *IEEE Transactions on Communications*, *IEEE Access* and *IEEE Open Journal of Vehicular Technology*. His research interests include multiple access, wireless AI and reconfigurable intelligent surface enabled communications. He has published more than 180 papers in IEEE journals with citations more than10,000 in Google scholar.

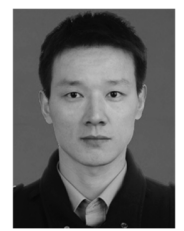

**He-Xiu Xu** (S'11, M'14, SM' 17) was born in China in 1985. He received the B.S. degree in radar engineering in June 2008, and the Ph.D. degree in electronic science and technology in June 2014 from Air Force Engineering University, Xi'an, China. He was a Visiting Scholar at the State Key Laboratory of Millimeter Waves, Southeast University, Nanjing, China, from February 2012 to January 2014, a Postdoctoral Fellow in the State Key Laboratory of Surface Physics, Fudan University, Shanghai, China from 2015 to 2017, and also a Visiting Scholar at the Department of Electrical and Computer Engineering, National University of Singapore, from December 2017 to December 2018.

Dr. Xu joined the Air Force Engineering University, in September 2014 and was promoted to an Associate Professor in 2016, and a full Professor in 2019. Currently, he is also a guest professor of Hengyang Normal University. He has published more than 150 peer-reviewed first-author and co-author journal papers in *Research*, *Nature Materials*, *Nature Photonics*, *Light: Sci. Appl.*, *Materials Today*, *Proceedings of IEEE*, *Advanced materials*, *Laser Photonics Rev.*, *ACS Photonics*, *Advanced Optical materials*, *Advanced Material Technologies*, *Photonics Research*, *IEEE Transactions on Microwave Theory and Techniques*, *IEEE Transaction on Antennas and Propagations, etc*. He has also published 4 Chinese books, 1 English book and 2 English book chapters. He served as an editor of *AEU Int. J. Electron. Commun* since 2014, a guest editor for Special Issue "Metamaterial Circuits and Antennas" of '*International Journal of RF and Microwave Computer-Aided Engineering*' in 2018, an associate editor for *Research*, and *IEEE Photonics Journal* since 2021, served as an editorial board member of several China journals such as *J. Infrared Millim. Wave*, *Journal of Electronics & Information Technology*, Transactions of Nanjing University of Aeronautics and Astronautics (TNUAA), and *Space Electronic Technology* since 2021, and served as an invited reviewer for more than 20 leading journals including *Light Science & Applications, etc*. He has owned 30 China patents and has given more than 50 invited talks. He served as a special session chair 25 times and a TPC co-chair/member more than 20 times in international conferences. His research interests include passive/active metamaterials/metasurfaces, and their applications to novel microwave functional devices and antennas.

Dr. Xu received the 8th China Youth Science and Technology Innovation Award in 2013. He won the best Excellent Doctoral Dissertation Award at Air Force Engineering University in 2014 and later received the Excellent Doctoral Dissertation Award from Military, Shaanxi Province, and Chinese Institute of Electronics (CIE). He has hosted several programs such as the National Science Foundation of China, National Defense Foundation of China, the Key Program of National Natural Science Foundation of Shaanxi Province, and First-Class General and Special Financial Grant from China Postdoctoral Science Foundation. He receives the URSI GASS Young Scientist Award, URSI AP-RASC Young Scientist Award, and URSI EMTS Young Scientist Award and ACES Young Scientist Award in 2019, URSI GASS Young Scientist Award in 2020 and Progress In Electromagnetics Research Symposium Young Scientist Award for SC2 and SC4 in 2019 and 2021, respectively. He was also granted several Scientific and Technological Progress Awards like China Patent Award, First Prize of Invention & Innovation Award from China Association of Inventions, Gold Award of National technological invention, three Special Prizes of Scientific and Technological Progress Award by Education Department of Shaanxi Provincial Government and Second Prize of Scientific and Technological Progress Award by CIE. He was awarded Young Scientist Nova from Shaanxi Technology Committee in 2016, a Young Talent from China Association for Science and Technology in 2017, outstanding scientific and technological worker of CIE in 2018, Technical Star Award by the Government of Xi'An, Outstanding Young Talent Award in Universities by the Education Department of Shaanxi Provincial Government and Young top talents of Shaanxi special talents supported by the Government of Shaanxi Province in 2019, and Cheung Kong Scholar in 2020. He was also the receipt of Excellent Technical Paper Award from China Association for Science and Technology in 2023, Outstanding Paper Awards from *Light Science & Applications* in 2019 and from *Research* in 2022, The 14th Excellent Scientific Paper Award from Shaanxi Government in 2019, Annual Excellent Research Paper Awards from CIE in conformal metasurface in 2020 and multiplexing metasurface in 2022, Excellent Paper Award in metasurface multifunctional devices at the 23th Annual Youth Conference of CIE in 2017, in helicity control of metasurface at PIER Symposium in 2018, and a co-recipient of three Best Paper Awards such as in bifunctional metasurfaces at the A3 Metamaterials Forum in 2017. He is ranked 2‰ in World Ranking of Scientists and Elsevier Global Highly Cited Scientist from 2020 to 2024. Dr. Xu is now also a Fellow of IET, a Fellow of RAeS, a Senior Member of CIE, and a committee member of antenna branch of CIE.